\documentclass[11pt]{article}
\usepackage{amssymb}
\usepackage{amsmath,amsfonts,amsthm,amssymb}
\usepackage{enumerate}
\usepackage{setspace}
\usepackage{lastpage}
\usepackage{extramarks}
\usepackage{algorithmic}
\usepackage{enumerate}
\usepackage{chngpage}
\usepackage{soul,color}
\usepackage{graphicx,float,wrapfig}

\theoremstyle{plain}
\newtheorem{thm}{Theorem}[section]
\newtheorem{lem}[thm]{Lemma}

\newtheorem{problem}[thm]{Problem}

\theoremstyle{definition}
\newtheorem{defn}[thm]{Definition}
\newtheorem{fact}[thm]{Fact}

\theoremstyle{remark}

\newcommand{\comment}[1]{}

\newcommand{\cC}{\mathcal C}
\newcommand{\cT}{\mathcal T}
\newcommand{\ccap}{{\text{cap}}}

\floatstyle{boxed}
\newfloat{Protocol}{h}{pro}

\def\proof{\noindent{\bf Proof:~}}

\newcommand{\Title}{Simulating Noisy Channel Interaction}
\newcommand{\Author}{Mark Braverman\thanks{Department of Computer Science, Princeton University, email: mbraverm@cs.princeton.edu. Research supported in part by an   an NSF CAREER award (CCF-1149888), a 
Turing Centenary Fellowship, a Packard Fellowship in Science and Engineering, and the Simons Collaboration on Algorithms and Geometry.} \and Jieming Mao\thanks{Department of Computer Science, Princeton University, email: jiemingm@princeton.edu}}
\newcommand{\enterProblemHeader}[1]{\nobreak\extramarks{#1}{#1 continued on next page\ldots}\nobreak%
                                    \nobreak\extramarks{#1 (continued)}{#1 continued on next page\ldots}\nobreak}%
\newcommand{\exitProblemHeader}[1]{\nobreak\extramarks{#1 (continued)}{#1 continued on next page\ldots}\nobreak%
                                   \nobreak\extramarks{#1}{}\nobreak}%

\newcommand{\homeworkProblemName}{}%
\newcounter{homeworkProblemCounter}%
  {\stepcounter{homeworkProblemCounter}%
   \renewcommand{\homeworkProblemName}{#1}%
   \section*{\homeworkProblemName}%
   \enterProblemHeader{\homeworkProblemName}}%
  {\exitProblemHeader{\homeworkProblemName}}%

\newcommand{\homeworkSectionName}{}%
\newlength{\homeworkSectionLabelLength}{}%
  {

   \renewcommand{\homeworkSectionName}{#1}%
   \settowidth{\homeworkSectionLabelLength}{\homeworkSectionName}%
   \addtolength{\homeworkSectionLabelLength}{0.25in}%
   \changetext{}{-\homeworkSectionLabelLength}{}{}{}%
   \subsection*{\homeworkSectionName}%
   \enterProblemHeader{\homeworkProblemName\ [\homeworkSectionName]}}%
  {\enterProblemHeader{\homeworkProblemName}%

   \changetext{}{+\homeworkSectionLabelLength}{}{}{}}%
   
\title{\Title}
\author{\Author}

\begin{document}
\begin{spacing}{1.1}
\maketitle \thispagestyle{empty}

\abstract{
We show that $T$ rounds of interaction over the binary symmetric channel $BSC_{1/2-\epsilon}$ with feedback can be 
simulated with $O(\epsilon^2 T)$ rounds of interaction over a noiseless channel. We also introduce a more general 
``energy cost'' model of interaction over a noisy channel. We show energy cost to be equivalent to external information 
complexity, which implies that our simulation results are unlikely to carry over to energy complexity. Our main technical 
innovation is a self-reduction from simulating a noisy channel to simulating a slightly-less-noisy channel, which may 
have other applications in the area of interactive compression.
} 

\section{Introduction}

Much of modern coding theory revolves around the following question: ``Given an imperfect (noisy) channel $\cC$, what is the best 
way of utilizing it to simulate noiseless communication?'' A key objective of Shannon's classical information theory \cite{Shannon48,cover2006elements}
was to answer this question. It turns out that for memoryless channels, the number of utilizations of $\cC$ needed to transmit 
$n$ bits of information scales as $n/\ccap(\cC)$, where $\ccap(\cC)$ is the {\em channel capacity} of $\cC$.

In this paper we consider the converse problem: 

\begin{problem}\label{pr:main}
Can a {\em noiseless} channel be {\bf effectively} utilized to simulate communication over a noisy channel $\cC$?
\end{problem}

We will focus entirely on binary channels with feedback --- i.e. channels transmitting bits $\in\{0,1\}$, where
the transmitting party gets to observe the (possibly corrupted) received bit --- although the 
results can likely be generalized to a broader class of channels. Note that as our discussion is about simulating a noisy channel with a noiseless one, the fact that the channel has feedback only makes such simulation more difficult. Most of our discussion will focus on the {\em binary symmetric channel} $\cC=BSC_a$, 
for noise $0\le a<1/2$. A bit $b$ transmitted over $BSC_a$ is received as $b\oplus b_{err}$, where $b_{err}\sim B_{a}$ is a Bernoulli random variable 
that causes the received bit to be flipped. It is well known that $\ccap(BSC_a) =1- H(a):=1 +a\log a + (1-a) \log (1-a)$. A particularly interesting 
regime in our context is when the noise level is very high: $a=1/2-\epsilon$. In this case $1-H(a) = \Theta(\epsilon^2)$. 

Of course, communication over a noisy channel 
can always be simulated by communication over a noiseless channel: the sender can simply apply the noise before transmitting
her bit to the receiver. However, one would like to simulate the communication {\em effectively}, only paying $O(\ccap(\cC)\cdot n)$
bits of communication to simulate $n$ utilizations of $\cC$. 

We will consider the problem in a general interactive setting, where $\cC$ is being used to conduct a general interactive protocol.
In the non-interactive setting, classical results from information theory show that up to factor $(1+\delta)$, with $\delta\rightarrow 0$ as $n\rightarrow\infty$, 
$n$ utilizations of $\cC$ can be simulated by $\sim \ccap(\cC)\cdot n$ utilizations of $\cC$, and vice-versa. What can one say about 
the interactive case?

Coding for interactive communication, i.e. encoding a noiseless protocol over a noisy channel (the converse problem to the one we are trying to solve) 
has received a substantial amount of attention recently. An early result by Schulman \cite{Schulman96} showed that good (constant-rate, constant-fraction-of-errors) codes exist in the interactive setting  even when the noise on the channel is adversarial. This work has since been recently improved in several directions, including error-tolerance and the code's computational efficiency \cite{braverman2011towards,brakerski2012efficient,ghaffari2013optimal1,ghaffari2013optimal2,braverman2014list}. Most relevant to our work is a result by Kol and Raz \cite{kol2013interactive}
showing a gap between interactive channel capacity and one-way channel capacity (interactive channel capacity is lower), once again giving an example 
of interactive coding theory being much more complicated than its one-way transmission counterpart. 

Problem~\ref{pr:main} can also be cast as a problem of {\em compressing interactive communication}. The general problem 
of compressing interactive communication arises in the context of information complexity and direct sum problems 
for randomized communication complexity \cite{ChakrabartiSWY01,BaryossefJKS04,BBCR,BravermanR11}. The (internal) information cost of a two-party protocol $\pi$ is the amount 
of information executing $\pi$ reveals to the parties about each other's inputs. 
In its full generality, interactive compression asks to simulate an information cost-$I$ protocol with $O(I)$ communication, and 
is equivalent to the strong direct sum problem in communication complexity \cite{BravermanR11}. Unfortunately, such strong interactive compression 
has recently been shown to be impossible \cite{ganor2014exponential}. A less ambitious goal is to compress $\pi$ to its external information cost 
$I^{ext}\ge I$. There are reasons to believe that compression to $O(I^{ext})$ communication is also impossible. For example, \cite{braverman2013hard} gives
a specific problem that is conjectured to provide such a separation. 

Communication over a noisy channel $BSC_{1/2-\epsilon}$ inherently reveals only $1-H(1/2-\epsilon)=\Theta(\epsilon^2)$ information 
to the observer in each round. Thus, a protocol $\pi$ that runs for $T$ rounds over such a channel has (both internal and external) information cost $O(\epsilon^2 T)$, 
although the {\em way} in which this information is limited round-by-round is highly structured. In this case, our first main result shows 
that compression with $O(1)$ multiplicative loss is possible:

\begin{thm} \label{thm:intromain1} {\em
(Theorem~\ref{thm:main1}, rephrased)} Any protocol $\pi$ running for $T$ rounds over  $BSC_{1/2-\epsilon}$ with feedback can be perfectly simulated
by a public-randomness protocol $\pi'$ running for $O(\epsilon^2 T)$ rounds in expectation over the noiseless channel $BSC_0$.
\end{thm}

Theorem~\ref{thm:intromain1} provides a new result on the cusp between information complexity theory and interactive coding theory. It shows that 
(up to a constant) interaction over a noisy channel can be simulated by interaction over noiseless channel, giving an affirmative action 
to Problem~\ref{pr:main} in this case. 

The compression proof of Theorem~\ref{thm:intromain1} relies crucially on the fact that errors on the channel remain the same throughout 
the communication. We consider the following strengthening of the error model: in each round, the party transmitting the next bit {\em chooses} the 
error rate $1/2-\epsilon$ of the next bit, while paying {\em energy cost} $EC$ of $\Theta(\epsilon^2)$. This model corresponds to a scenario 
where the party gets to modulate its transmission power in a way that affects the noise level (and thus the channel capacity) of the transmission.
While we chose $\epsilon^2$ because it captures the channel capacity for the selected $\epsilon$, this expression is known to capture
actual energy-capacity tradeoffs in high-noise wireless scenarios (see e.g. \cite{TV05}). 
We show that thus defined energy complexity is actually equivalent to the external information complexity:

\begin{thm} \label{thm:intromain2} {\em
(Theorems~\ref{eclb} and \ref{ecub}, rephrased)} For any
 protocol $\pi$ over a variable-noise $BSC$ with feedback and any distribution $\mu$ over inputs, there is a protocol $\phi$ over a noiseless channel, such that the external information cost of $\phi$ is $O( EC_{\mu}(\pi)) $ and $\phi$ simulates $\pi$. Conversely, any $\phi$ with external 
information cost $I^{ext}$ can be simulated by a $\pi$ over  a variable-noise $BSC$ with feedback with $ EC_{\mu}(\pi) = O(I^{ext}+\epsilon)$ for 
any $\epsilon>0$. 
\end{thm}

Theorem~\ref{thm:intromain2} implies that the analogue of Theorem~\ref{thm:intromain1} is unlikely to hold for the more general variable-error model, 
since it is believed that one cannot compress a general interactive protocol $\pi$ to $O(IC^{ext}(\pi))$. We note that the strongest known 
compression result is of the form $O(IC^{ext}(\pi) \cdot (\log |\pi|)^{O(1)})$ \cite{BBCR}, where $|\pi|$ is the number of bits communicated by $\pi$. 

We believe that techniques involved in proving Theorem~\ref{thm:intromain1} (discussed below) have the potential to be helpful 
in compressing interactive communication. While we know by \cite{ganor2014exponential} that compressing $\pi$ all the way down to $IC(\pi)$ is impossible, 
one can hope to beat the currently best compression scheme of $\tilde{O}(\sqrt{IC(\pi)\cdot |\pi|})$ of \cite{BBCR}. Specifically, to the best
of our knowledge, the recursive approach we describe below has not appeared in past works in either the Information Theory or the Theoretical 
Computer Science literature. 

\subsection{Techniques and proof overview of Theorem~\ref{thm:intromain1}}
\label{sec:intuition}

In this section we briefly discuss the technical contributions of this paper. We will mainly focus on the techniques in 
the proof of Theorem~\ref{thm:intromain1}: while the proof of Theorem~\ref{thm:intromain2} requires care and work, it 
does build on existing techniques from past works in the area, such as \cite{braverman2013information}. 

Recall that to prove Theorem~\ref{thm:intromain1} we need to take a protocol $\pi$ that runs for $T$ steps over $BSC_{1/2-\epsilon}$, and 
simulate it using a protocol $\phi$ that runs for $O(\epsilon^2 T)$ steps over the noiseless channel $BSC_0$. A natural approach 
is to break $\pi$ into ``chunks'' of $\Theta(1/\epsilon^2)$ communication each, and to try and simulate each chunk using $O(1)$ communication. 
Let $\pi'$ denote a sub-protocol of $\pi$ of $\gamma=1/\epsilon^2$ rounds we are trying to simulate. There is a natural way 
to identify transcripts of $\pi'$ with leafs of a binary tree $\cT$ of depth $\gamma$. Each leaf $\ell$ corresponds to a transcript that contains 
$0\le m \le \gamma$ mistakes. The goal of the parties (Alice and Bob) is to sample each $\ell$ with its correct probability 
$p_\ell:=(1/2-\epsilon)^m (1/2+\epsilon)^{\gamma-m}$. Note that for a given $\ell$, Alice and Bob do not know $m$. Rather, since each of them 
only knows what part of his or her messages were corrupted, Alice and Bob know two numbers $m_x$ and $m_y$, respectively, such 
that $m=m_x+m_y$. 

Following past works, Alice and Bob can try to first jointly sample a leaf $\ell$ and then use rejection sampling to make sure that each $\ell$ is 
selected with probability proportional to $p_\ell$. Since the joint sampling happens without any communication, we select each leaf with 
probability $2^{-\gamma}$. Note that under such a procedure no leaf ever gets selected with probability $>2^{-\gamma}$, thus if we want to accommodate
leafs with $p_\ell>2^{-\gamma}$ we should select each leaf with probability $p_\ell/M$ for a constant $M>1$. Note that this means that each round 
will succeed with probability $\sim 1/M$, and thus we can only afford $M=O(1)$ a large constant. This will allow Alice and Bob to sample {\em most}
but not {\em all} leafs correctly. Note that the probability of the most likely leaf in $\cT$ is $2^{-\gamma} \cdot (1+2 \epsilon)^\gamma \sim 
2^{-\gamma} \cdot e^{2/\epsilon} \gg 2^{-\gamma}$, and our rejection sampling approach is bound to fail here by badly under-sampling this leaf. 

A (partial) solution to the problem above is to choose $\gamma$ slightly smaller than $1/\epsilon^2$ (e.g. $1/(\epsilon^2 \log |\pi|)$), and 
just ignore leafs for which the ratio exceeds $M$. This is the approach employed in \cite{BBCR} to compress to external information cost. One 
can show that at each round we add small (e.g. $<1/|\pi|^2$) statistical error, and thus the simulation (mostly) works. This approach is unsuitable for us here for two reasons. Firstly, we would like to have a perfect simulation that does not incur any error. Secondly, in order to get
a $O(1)$-bit simulation of $\pi'$ we cannot afford the depth of $\cT$ to be $o(1/\epsilon^2)$. 

Instead, we adopt a recursive approach. We begin the simulation of $\pi'$ by tossing (a properly biased) coin, and deciding whether we will 
be looking for a ``high-error'' or a ``low-error'' leaf, where the threshold distinguishing ``high'' and ``low'' is chosen appropriately (note that the ``low-error'' leafs are the ones getting under-counted by the rejection
sampling protocol). If we are looking for ``high-error'' nodes, then rejection sampling with an appropriate constant $M>1$ as described above will 
work well. What should we do about a ``low-error'' leaf? We would like to sample such a leaf $\ell$ with probability exceeding $p_\ell$, since we are only trying to sample it conditioned on entering the ``low-error'' regime. To get such a sampling for the low error regime we just  
simulate $\pi'$, but over $BSC_{1/2-2\epsilon}$ instead of $BSC_{1/2-\epsilon}$! We use induction to claim such a sampling is possible (note that when $\epsilon=\Theta(1)$ simulation is trivial since $|\pi'|=1/\epsilon^2 = O(1)$). Simulating $\pi'$ over a lower noise channel $BSC_{1/2-2\epsilon}$ has the effect of ``punishing'' high-error leafs (we don't care about those since they get sampled in the high error regime), and ``rewarding''
low-error leafs, which are the ones we would like to focus on. For example, the most likely no-errors leafs is approximately $e^{2/\epsilon}$ times 
more likely under $BSC_{1/2-2\epsilon}$ than under $BSC_{1/2-\epsilon}$. Of course, simulating $\pi'$ over $BSC_{1/2-2\epsilon}$  is more expensive 
than over $BSC_{1/2-\epsilon}$ --- $\approx 4$ times more expensive as $(1/\epsilon)^2\cdot (2\epsilon)^2= 4$ --- but as long as the low-error regime 
is invoked $<1/4$ of the time, the total communication converges and remains $O(1)$ in expectation. 

As the problem of sampling ``low-error'' nodes is the main difficulty in the general compression of interactive communication, we hope that the 
strategy above will be helpful in addressing this more general problem.

\subsection{Techniques and proof overview of Theorem~\ref{thm:intromain2}} 
In this section we give a proof overview of Theorem~\ref{thm:intromain2}. Theorem~\ref{thm:intromain2} has two parts. We will discuss them separately.

Recall that the first part of Theorem~\ref{thm:intromain2} shows that for any protocol $\pi$ over a variable-noise BSC with feedback and any distribution $\mu$ over inputs, we can construct a protocol $\phi$ over a noiseless channel, such that the external information cost of $\phi$ is $O(EC_{\mu}(\pi))$ and $\phi$ simulates $\pi$. The proof of this part of Theorem~\ref{thm:intromain2} is straightforward. For each bit $b$ transmitted over $BSC_p$ in protocol $\pi$, the transmitter sends $b \oplus B_p$ to the receiver over a noiseless channel in $\phi$. The analysis of external information cost of $\phi$ follows the standard information-theoretic argument which first converts the information cost into  the sum of the divergence between the true probability and the prior information and then bounds the divergence by the energy cost.

The second part of Theorem~\ref{thm:intromain2} shows that for any protocol $\phi$ over a noiseless channel, we can construct a protocol $\pi$ over a variable-noise BSC with feedback, such that $EC_{\mu}(\pi) = O(IC^{ext}(\phi) + \epsilon)$ for any $\epsilon >0$. Our approach considers protocol $\phi$ bit by bit. For each transmitted bit in $\phi$, let's assume the transmitter wants to send this bit as $B_p$ and both the transmitter and the receiver have prior information $B_q$. Then the external information cost of this bit is $D(p\|q)$. This divergence is the budget for the energy cost of the corresponding part in $\pi$. 

The general protocol we used in this proof to send $B_p$ with prior $B_q$ and energy cost $D(p\|q)$ does a biased random walk on points $0,\frac{1}{2n},...,\frac{2n-1}{2n},1$. Here $n$ is some previously fixed integer. For this biased random walk, the transmitter and the receiver agree to start at some point closest to $q$. The transmitter starts to send bits over some chosen binary symmetric channels and they move left or right according to received bits. They stop this biased random walk when they reach either 0 or 1, and they pick the sampled bit as the stop position. The main technique used in this biased random walk is Lemma~\ref{brw}. This lemma shows that if we do biased random walk on points $0,1,...a-1,a,a+1,...a+b-1,a+b$, starting at point $a$ and $a \geq b$, then the transmitter only needs to spend a constant energy cost to always end at point $a+b$. Directly from this lemma, the transmitter can go from point $q$ to point $q\cdot 2^t$ with energy cost $O(t)$. 

Unfortunately, under this biased random walk framework, it is difficult to design an integral protocol for all kinds of $p$ and $q$. So for different values of $p$ and $q$, our approach uses different lower bounds of $D(p\|q)$ as the budget for energy cost. In each case, the transmitter will use Lemma~\ref{brw} differently to meet the lower bounds of $D(p\|q)$. Table 1 shows the lower bounds of $D(p\|q)$ used in different cases. 

The $\epsilon$ in the energy cost comes from the fact that $q$ might not be a point where we do random walk (i.e. $\frac{i}{2n}$). So we will start with a point closest to $q$, and this approximation will make the energy cost increase by $O(\epsilon)$. In fact, this $\epsilon$ equals to $\frac{1}{2n}$. As increasing $n$ will not make the energy cost increase, we can make this $\epsilon$ arbitrarily small.

\begin{table}
\centering
\caption{Divergence lower bound}
\begin{tabular}{|c|c|}
\hline
Cases & Lower bounds of $D(p\|q)$\\
\hline
\hline
$0\leq p \leq 2q$ & $\Omega(\frac{(p-q)^2}{q})$\\
\hline
$2q < p < 0.02$, $q< 0.01$ & $\Omega(p\log \frac{p}{q})$\\
\hline
$2q < p$, $q \geq 0.01$ & $\Omega(1)$\\
\hline
$p \geq 0.02$, $q < 0.01$ & $\Omega(\log \frac{1}{q})$\\
\hline
\end{tabular}
\end{table}

%
%

\section{Preliminaries}
\subsection{Communication Complexity}
In the two-party communication model, Alice and Bob want to jointly compute a function $f: \mathcal{X} \times \mathcal{Y} \rightarrow \mathcal{Z}$. Alice is only given input $x \in \mathcal{X}$ and Bob is only given input $y \in \mathcal{Y}$. 
In this paper, we consider the public coin model, which means that Alice and Bob have access to the shared randomness. In order to compute function $f$, they have to communicate with each other following a protocol $\pi$ which specifies when the communication is over, who sends the next bit if the communication is not over, and the function of each transmitted bit given the history, the input of the person who sends this bit and the shared randomness. The transcript of a protocol is a concatenation of all bits exchanged. 
\begin{defn} 
The \emph{communication complexity} of a public coin protocol $\pi$, denoted by $CC(\pi)$, is defined as the maximum number of bits exchanged on the worst input. 
\end{defn}
\begin{defn} 
The \emph{average communication complexity} of a public coin protocol $\pi$, denoted by $\overline{CC}(\pi)$, is defined as the maximum expected number of bits exchanged over the randomness of the protocol on the worst input. 
\end{defn}

\begin{defn}
We will say that a protocol $\phi$ over a noiseless channel \emph{simulates} a protocol $\pi$ over a noisy channel if there is a deterministic function $g$ such that $g(\Phi(x,y,R^{\phi},R^{\phi}_A,R^{\phi}_B))$ is equal in distribution to $\Pi(x,y,R^{\pi}, R^{\pi}_A, R^{\pi}_B, R^c)$ for all $x$ and $y$. Here $R^{\phi}$ and $R^{\pi}$ are the public randomness used in protocol $\phi$ and $\pi$. $R^{\phi}_A$, $R^{\phi}_B$, $R^{\pi}_A$, $R^{\pi}_B$ are the private randomness used in protocol $\phi$ and $\pi$. $R^c$ is the randomness for the noisy channel. $\Pi$ and $\Phi$ are random variables for transcripts of protocols $\pi$ and $\phi$.
\end{defn}

\begin{defn}
We will say that a  protocol $\pi$ over a noisy channel \emph{simulates} a protocol $\phi$ over a noiseless channel if there is a deterministic function $g$ such that $g(\Pi(x,y,R^{\pi}, R^{\pi}_A, R^{\pi}_B, R^c))$ is equal in distribution to $\Phi(x,y,R^{\phi}, R^{\phi}_A, R^{\phi}_B)$ for all $x$ and $y$. Here $R^{\phi}$ and $R^{\pi}$ are the public randomness used in protocol $\phi$ and $\pi$. $R^{\phi}_A$, $R^{\phi}_B$, $R^{\pi}_A$, $R^{\pi}_B$ are the private randomness used in protocol $\phi$ and $\pi$. $R^c$ is the randomness for the noisy channel. $\Pi$ and $\Phi$ are random variables for transcripts of protocols $\pi$ and $\phi$.
\end{defn}

Additional definitions and results in basic communication complexity can be found in \cite{KushilevitzN97}.

\subsection{Binary Symmetric Channel and Energy Cost}
\begin{defn}
The \emph{binary symmetric channel} with crossover probability $p$ ($0\leq p \leq \frac{1}{2}$), denoted by $BSC_p$, is defined as a communication channel such that each bit sent by the transmitter is flipped with probability $p$ when received by the receiver. 
\end{defn}

\begin{defn}
The $BSC_p$ with \emph{feedback} is defined as the $BSC_p$ such that the transmitter also gets the (potentially flipped) bit which the receiver receives. 
\end{defn}
In this paper, we consider two kinds of  two-party communication protocols over binary symmetric channels. One is that the crossover probability of the channel is fixed during the whole protocol. The other is that the transmitter can choose the crossover probability of the binary symmetric channel for each transmitted bit and the receiver does not know the crossover probability. For protocols in these two models, we can still define the communication complexity as the maximum number of bits exchanged. However, the following definition of energy cost is more close to the sense of information exchanged in the protocol. 

\begin{defn}
If the transmitter sends one bit over $BSC_p$ with feedback, the \emph{energy cost} of this bit is defined as $4(p-\frac{1}{2})^2$. The \emph{energy cost} of a protocol $\pi$ over binary symmetric channels(may have different crossover probabilities) with feedback, denoted by $EC(\pi)$, is defined as the maximum expected sum of \emph{energy cost} of each transmitted bit of $\pi$ over the randomness of the protocol on the worst input. 
\end{defn}

\begin{defn}
Given a distribution $\mu$ on inputs $X,Y$, the \emph{distributional energy cost}, denoted by $EC_{\mu}(\pi)$, is defined as the expected sum of energy cost of each transmitted bit of $\pi$ over input distribution $\mu$ and the randomness of the protocol.
\end{defn}

\subsection{Information Theory and Information Cost }

More definitions and results from basic information theory can be found in \cite{cover2006elements}. All the $\log$s in this 
paper are base $2$. 

\begin{defn}
The \emph{entropy} of a random variable $X$, denoted by $H(x)$, is defined as $H(X) = \sum_x Pr[X = x] \log(1/Pr[X = x])$. 
\end{defn}
If $X$ is drawn from Bernoulli distributions $B_p$, we use $h(p) = -(p\log p + (1-p)(\log(1-p))$ to denote $H(X)$. 
\begin{defn}
The \emph{conditional entropy} of random variable $X$ conditioned on random variable $Y$ is defined as $H(X|Y) = \mathbb{E}_y[H(X|Y = y)]$. 
\end{defn}
\begin{fact}
$H(XY) = H(X) + H(Y|X)$. 
\end{fact}
\begin{defn}
The \emph{mutual information} between two random variables $X$ and $Y$ is defined as $I(X;Y) = H(X) - H(X|Y) = H(Y) - H(Y|X)$. 
\end{defn}
\begin{defn}
The \emph{conditional mutual information} between $X$ and $Y$ given $Z$ is defined as $I(X;Y|Z) = H(X|Z) - H(X|YZ) = H(Y|Z) - H(Y|XZ)$. 
\end{defn}
\begin{fact}\label{cr}
Let $X_1,X_2,Y,Z$ be random variables, we have $I(X_1X_2;Y|Z) = I(X_1;Y|Z) + I(X_2;Y|X_1Z)$.
\end{fact}
\begin{defn}
The \emph{Kullback-Leibler divergence} between two random variables $X$ and $Y$ is defined as $D(X\| Y) = \sum_x Pr[X = x] \log(Pr[X = x] / Pr[Y = x])$. 
\end{defn}
If $X$ and $Y$ are drawn from Bernoulli distribution $B_p$ and $B_q$, we use $D(p\| q)$ as an abbreviation of $D(X\| Y)$. 
\begin{fact}\label{div}
Let $X,Y,Z$ be random variables, we have $I(X;Y|Z) = \mathbb{E}_{x,z}[D((Y|X = x, Z=z)\|(Y|Z=z))]$.
\end{fact}
\begin{fact}\label{ine}
Let $X,Y$ be random variables, 
\[
\sum_x \frac{|Pr[X = x] - Pr[Y = x]|^2}{2\max(Pr[X = x], Pr[Y=x])} \leq \ln(2) \cdot D(X\|Y) \leq \sum_x \frac{|Pr[X = x] - Pr[Y = x]|^2}{Pr[Y=x]}
\]
\end{fact}

\proof

For notation convenience, let $p(x) = Pr[X = x]$ and $q(x) = Pr[Y = x]$. Let's first prove the right-hand side. 
\begin{eqnarray*}
\ln(2) \cdot  D(X\| Y) &=& \sum_x p(x) \ln (\frac{p(x)}{q(x)}) \\
&\leq& \ln (\sum_x \frac{p(x)^2}{q(x)}) ~~~~\text{(by concavity of $\ln(z)$)}\\
&\leq& \sum_x \frac{p(x)^2}{q(x)} - 1\\
&=& \sum_x \frac{(p(x) - q(x))^2}{q(x)}\\
\end{eqnarray*}

For the left-hand side, consider any convex function $f$ such that $f''(x) \geq m > 0$ for all $x \in [a,b]$. By strong convexity, for $x,y \in [a,b]$, we have 
\[
f(y) \geq f(x) + f'(x)(y-x) + \frac{m(y-x)^2}{2}.
\]
Let $f(x) = x \ln x$. For $x \in [a,b]$, we have $f''(x) \geq \frac{1}{b}$. Therefore,
\[
a\ln a \geq b \ln b + (a-b)(1+\ln b) + \frac{(a-b)^2}{2b}.
\]
and then
\[
a \ln(\frac{a}{b}) \geq (a-b) +  \frac{(a-b)^2}{2b}. 
\]
Similarly, we have
\[
b \ln(\frac{b}{a}) \geq (b-a) +  \frac{(a-b)^2}{2b}. 
\]
Thus
\begin{eqnarray*}
\ln(2) \cdot  D(X\| Y) &=& \sum_x p(x) \ln (\frac{p(x)}{q(x)}) \\
&\geq &\sum_x [p(x) - q(x) + \frac{(p(x) - q(x))^2}{2\max\{p(x),q(x)\}}]\\
&=& \sum_x\frac{(p(x) - q(x))^2}{2\max\{p(x),q(x)\}}\\
\end{eqnarray*}
\qed

Finally, we define the (external) information cost of a protocol. 

\begin{defn}
Given a distribution $\mu$ on inputs $X$,$Y$, and a public coin protocol $\pi$, the \emph{external information cost} is defined as $IC_{\mu}^{ext}(\pi) = I(XY; \Pi)$, where $\Pi = \Pi(X,Y,R)$ is the random variable denoting the transcript and public randomness of the protocol and $R$ is the public randomness. 
\end{defn}

\section{Simulating the noise channel using the noiseless channel}
\begin{thm} \label{thm:main1}
For every deterministic protocol $\pi$ over $BSC_{1/2 - \epsilon}$ with feedback, there exists a public coin protocol $\phi$  over noiseless channel such that $\phi$ simulates $\pi$ and 
\[
\overline{CC}(\phi) \leq \alpha \cdot \lceil \epsilon^2 \cdot  2CC(\pi) \rceil.
\]
Here $\alpha$ is a constant and equals to $\max(\frac{1}{\beta^2},50t^2 + 10)$ where $t = e^6$ and $\beta$ is a constant to be determined in the proof. 
\end{thm}

\paragraph{Proof overview.} The proof follows the intuition outlined in Section~\ref{sec:intuition}. In the language of the overview, 
protocol $\phi_{v, \gamma, 1/2 - \epsilon}$, which is the main protocol simulating $\gamma$ layers starting from node $v$ in 
the protocol tree, decides whether to call $\phi^0_{v, \gamma, 1/2 - \epsilon}$ or $\phi^1_{v, \gamma, 1/2 - \epsilon}$.
$\phi^1_{v, \gamma, 1/2 - \epsilon}$ takes care of the ``high-error'' regime case, and is executed using rejection sampling. 
$\phi^0_{v, \gamma, 1/2 - \epsilon}$ takes care of the ``low-error'' regime case, and uses a recursive call to the execution of $\pi'$ over
$BSC_{1/2-2\epsilon}$, followed by rejection sampling to make probabilities align perfectly. 

One technical detail which we omitted from the the intuitive description but that plays an important role in the protocols is
the $threshold_{\theta, v, w, D}$ function. In order to be able to perform rejection sampling starting from a node $v$, we need to know whether a given node $w$ located $\gamma$ layers below $v$ has more errors than the ``high-error'' threshold $\theta$ or less. This depends on whether 
the number of mistakes $m_x+m_y$ along the path from $v$ to $w$ exceeds $\theta$ or not. Unfortunately, only Alice knows $m_x$ and only Bob knows 
$m_y$, and exchanging these values is prohibitively expensive: it would cost $\Theta(\log \gamma)$ bits of communication, whereas we can only 
afford $O(1)$ communication to perform this operation. Luckily, for nodes sampled from $D$, if the distribution of $(m_x,m_y)$ is a product distribution (it is in our case), we are able to give an expected $O(1)$ protocol for the problem. In addition to answering whether $m_x+m_y>\theta$, the protocol $threshold_{\theta, v, w, D}$ outputs a pair of ``witnesses'' $(\theta_x,\theta_y)$ such that $\theta_x+\theta_y = \theta$ that work as follows: if $m_x+m_y\le \theta$, then $m_x\le \theta_x$ and $m_y\le \theta_y$; if $m_x+m_y>\theta$, then $m_x\ge \theta_x$ and $m_y\ge \theta_y$. These witnesses are then used 
by Alice and Bob when performing rejection sampling. 

\medskip

\proof
First we change $\pi$ to be the protocol that Alice and Bob send messages alternatively. This modification will increase $CC(\pi)$ by at most a multiplicative factor of $2$.

 Now we consider the easy case when $\epsilon \geq \beta$. In this case, we just make $\phi$ to be the direct simulation of $\pi$. That is, if in protocol $\pi$ Alice has to send a bit $b$, then in protocol $\phi$, Alice sends the same bit $b$, and both Alice and Bob use public randomness to generate $b' \sim B_{1/2 -\epsilon}$ and pretends the receiving bit to be $b \oplus b'$. In this way, the bit Bob receives in $\phi$ will have the same distribution as the bit Bob receives in $\pi$. When Bob sends a message in $\pi$, we do the same modification in $\phi$. Therefore $\phi$ simulates $\pi$ and $\overline{CC}(\phi)\leq 2CC(\pi) \leq \frac{1}{\beta^2}\cdot \epsilon^2 \cdot 2CC(\pi) \leq \alpha \cdot \epsilon^2 \cdot  2CC(\pi)$. 

Now we prove this theorem by induction on the crossover probability for the case when $\epsilon < \beta$, showing that the theorem for $2\epsilon$ 
implies it for $\epsilon$. We construct $\phi$ by compressing $\gamma = \frac{1}{\epsilon^2}$ communication bits of $\pi$ over $BSC_{1/2-\epsilon}$ into a protocol over a noiseless channel with constant communication bits. For each step of the compression, we consider $\gamma$ bits of $\pi$ as a protocol tree with root node $v$ and depth $\gamma$. In order to simulate this protocol tree, we only have to sample the leaf nodes with the same probabilities sampled from the protocol tree. The following protocols show how to do this. The main protocol is  protocol $\phi_{v, \gamma, 1/2 - \epsilon}$. For notation convenience, we define $m_x(v,w)$ to be the number of errors Alice makes from node $v$ to node $w$ on the protocol tree of $\pi$, $m_y(v,w)$ to be the number of errors Bob from node $v$ to node $w$ on the protocol tree of $\pi$, and $m(v,w) = m_x(v,w) + m_y(v,w)$. 
\begin{Protocol}
\caption{Protocol $\phi_{v, \gamma, 1/2 - \epsilon}$}
\begin{enumerate}
\item Let $\theta = \gamma \cdot (1/2 - 3\epsilon)$. Both players use public randomness to sample a bit $b$ from Bernoulli distribution $B_p$, where $p = \sum_{i = 0}^{i \leq \theta}(1/2-\epsilon)^i (1/2+\epsilon)^{\gamma - i} \binom{\gamma}{i}$. 
\item Run $\phi^b_{v, \gamma, 1/2 - \epsilon}$.
\end{enumerate}
\end{Protocol}

\begin{Protocol}
\caption{Protocol $\phi^0_{v, \gamma, 1/2 - \epsilon}$}
\begin{enumerate}
\item Alice and Bob pretend the crossover probability of the protocol tree is $1/2 - 2\epsilon$ and run $\phi_{v,\gamma, 1/2 - 2\epsilon}$ to sample a leaf node $w$. 
\item Alice and Bob run $threshold_{\theta, v, w,D}$. Here $D$ is the distribution from which $w$ is sampled, which satisfies $Pr_{w \sim D}[w] = (1/2 - 2\epsilon)^{m(v,w)}(1/2 + 2\epsilon)^{\gamma - m(v,w)}$. If the result if 1, they repeat this protocol. 
\item Alice samples a bit $b_x$ which is 1 with probability 
\[
\frac{(1/2 - \epsilon)^{m_x(v,w) - \theta_x}(1/2+\epsilon)^{- m_x(v,w) + \theta_x}}{(1/2 - 2\epsilon)^{m_x(v,w) - \theta_x}(1/2+2\epsilon)^{ - m_x(v,w) + \theta_x}},
\]
and sends this bit to Bob. Here $\theta_x$ gets its value from the previous run of $threshold_{\theta, v, w,D}$.
\item Bob samples a bit $b_y$ which is 1 with probability 
\[
\frac{(1/2 - \epsilon)^{m_y(v,w) - \theta_y}(1/2+\epsilon)^{- m_y(v,w) + \theta_y}}{(1/2 - 2\epsilon)^{m_y(v,w) - \theta_y}(1/2+2\epsilon)^{ - m_y(v,w) + \theta_y}},
\]
and sends this bit to Alice. Here $\theta_y$ gets its value from the previous run of $threshold_{\theta, v, w,D}$.
\item If both $b_x$ and $b_y$ are 1, they accept $w$. Otherwise they repeat this protocol.
\end{enumerate}
\end{Protocol}

\begin{Protocol}
\caption{Protocol $\phi^1_{v, \gamma, 1/2 - \epsilon}$}
\begin{enumerate}
\item Alice and Bob use public randomness to sample a leaf node $w$ of the protocol tree rooted at $v$ with depth $\gamma$ from the uniform distribution. Therefore each leaf node is sampled with probability $2^{-\gamma}$. 
\item Alice and Bob run $threshold_{\theta, v, w, D}$. Here $D$ is the uniform distribution on all leaf nodes. If the result is 0, they repeat this protocol.
\item Let $t =e^6$. 
\item Alice samples a bit $b_x$ which is 1 with probability 
\[
\frac{(1/2-\epsilon)^{m_x(v,w)}(1/2 +\epsilon)^{\gamma/2 -m_x(v,w)}}{t\cdot (\frac{1/2 - \epsilon}{1/2 +\epsilon})^{\theta_x - \theta / 2} \cdot 2^{-\gamma/2}},
\]
and sends this bit to Bob. Here $\theta_x$ gets its value from the previous run of $threshold_{\theta, v, w,D}$.
\item Bob samples a bit $b_y$ which is 1 with probability 
\[
\frac{(1/2-\epsilon)^{m_y(v,w)}(1/2 +\epsilon)^{\gamma/2 -m_y(v,w)}}{t\cdot (\frac{1/2 - \epsilon}{1/2 +\epsilon})^{\theta_y - \theta / 2} \cdot 2^{-\gamma/2}},
\]
and sends this bit to Alice. Here $\theta_y$ gets its value from the previous run of $threshold_{\theta, v, w,D}$.
\item If both $b_x$ and $b_y$ are 1, they accept $w$. Otherwise they repeat this protocol.
\end{enumerate}
\end{Protocol}

\begin{Protocol}
\caption{Protocol $threshold_{\theta, v, w, D}$}
\begin{enumerate}
\item Both players find integer $\xi$ such that $\Pr_{u\sim D} [m_x(v,u) \leq \xi - 1] \leq Pr_{u\sim D}[m_y(v,u) \leq \theta - \xi]$ and $Pr_{u\sim D}[m_x(v,u) \leq \xi] \geq Pr_{u\sim D}[m_y(v,u)\leq \theta - \xi - 1]$. 
\item Alice outputs a bit $b_1$ which is 1 if $m_x(v,w) = \xi$ and 0 otherwise. 
\item Alice outputs a bit $b_2$ which is 1 if $m_x(v,w) > \xi$ and 0 otherwise.
\item Bob outputs a bit $b_3$ which is 1 if $m_y(v,u) = \theta - \xi$ and 0 otherwise.
\item Bob outputs a bit $b_4$ which is 1 if $m_y(v,w) > \theta-\xi$ and 0 otherwise.
\item If $b_1 = 1$, the protocol returns $b_4$ and sets $\theta_x = \xi$ and $\theta_y = \theta - \xi$. 
\item If $b_1 = 0$ and $b_3 = 1$, the protocol returns $b_2$ and sets $\theta_x = \xi$ and $\theta_y = \theta-\xi$. 
\item If $b_1 = 0$,$b_3 = 0$,$b_2 = b_4$, the protocol returns $b_x$, and sets $\theta_x = \xi$ and $\theta_y = \theta - \xi$. 
\item If $b_1 = 0$,$b_3 = 0$,$b_2 = 1$,$b_4 = 0$, the protocol returns $threshold_{\theta, v, w, D|m_x(v,u) > \xi,m_y(v,u) < \theta -\xi}$.
\item If $b_1 = 0$,$b_3 = 0$,$b_2 = 0$,$b_4 = 1$, the protocol returns $threshold_{\theta, v, w, D|m_x(v,u) < \xi,m_y(v,u) > \theta -\xi}$.
\end{enumerate}
\end{Protocol}

Now let's intuitively understand how this set of protocols work. The set of protocols first divide the leaf nodes into two sets: $\{u|m(v,u) \leq \theta\}$ and $\{u|m(v,u) > \theta\}$. Since for each leaf node $u$, the probability that $u$ is sampled is $(1/2 - \epsilon)^{m(v,u)}(1/2 + \epsilon)^{\gamma - m(v,u)}$, the probability that nodes in the first set are sampled is exactly $p$. Then the protocol uses $\phi^0_{v, \gamma, 1/2 - \epsilon}$ to sample a node in the first set and $\phi^1_{v, \gamma, 1/2 - \epsilon}$ to sample a node in the second set. $\phi^0_{v, \gamma, 1/2 - \epsilon}$ uses the induction result of sampling a node with smaller crossover probability and $\phi^1_{v, \gamma, 1/2 - \epsilon}$ uses rejection sampling to sample a node in the second set. Both of these two protocols use protocol $threshold$ to determine whether the sampled node $w$ has  $m(v,w)$ greater than $\theta$ or not. 

Now let's analyze these protocols.

\textbf{Analysis of $threshold_{\theta, v, w, D}$: } This protocol's goal is to decide whether $m_x(v,w) + m_y(v,w) \leq \theta$ or not using only constant number of communication bits in expectation. This protocol will also make Alice and Bob get $\theta_x$ and $\theta_y$ which satisfy the following conditions:
\begin{itemize}
\item $\theta_x + \theta_y = \theta$.
\item If $m_x(v,w) + m_y(v,w) \leq \theta$, then $m_x(v,w) \leq \theta_x$ and $m_y(v,w) \leq \theta_y$. 
\item If $m_x(v,w) + m_y(v,w) > \theta$, then $m_x(v,w) \geq \theta_x$ and $m_y(v,w) \geq \theta_y$. 
\end{itemize}

The input distribution $D$ is the distribution where $w$ is sampled. This protocol only works for product distributions. More precisely, this protocol works when $m_x(v,w)$ has the same distribution as $m_x(v,w)$ given $m_y(v,w)$ to be any value and $m_y(v,w)$ has the same distribution as $m_y(v,w)$ given $m_x(v,w)$ to be any value. To analyze this protocol, we first have to make sure that in the first step of this protocol, the integer $\xi$ exists. Consider the following two conditions:
\begin{itemize}
\item $Pr[m_x(v,w) \leq \zeta] \leq Pr[m_y(v,w) \leq \theta - \zeta - 1]$.
\item $Pr[m_x(v,w) \leq \zeta] \geq Pr[m_y(v,w) \leq \theta - \zeta - 1]$.
\end{itemize}
For any integer $\zeta$, at least one of these two conditions will be satisfied. Also, we know that when $\zeta = -1$, the first condition is satisfied and when $\zeta = \theta$, the second condition is satisfied. So if when $\zeta = -1$, the second condition is also satisfied, we just have to pick $\xi = -1$. Otherwise, we can find some $\zeta$ between $-1$ and $\theta$ such that it violates the second condition and $\zeta+1$ satisfies the second condition. Then we just have to pick $\xi = \zeta + 1$. 

Finally let's analyze the communication cost of this protocol. Let $p_1 = Pr_{u\sim D}[m_x(v,u) \leq \xi - 1]$, $p_2 =Pr_{u \sim D}[m_x(v,u) \leq \xi]$,  $q_1=Pr_{u\sim D}[m_y(v,u) \leq \theta - \xi] $ and $q_2 = Pr_{u\sim D}[m_y(v,u) \leq \theta -\xi - 1]$. The probability that this protocol recursively calls itself at step 9 is $(1-p_2)q_2 \leq (1-p_2)p_2 \leq \frac{1}{4}$. The probability that this protocol recursively calls itself at step 10 is $p_1(1-q_1) \leq q_1(1-q_1) \leq \frac{1}{4}$.  Therefore, the probability that this protocol ends in one round is at least $\frac{1}{2}$. In expectation, Alice and Bob will communicate $4 \times 2 = 8$ bits running this protocol. In addition, if $D$ is a product distribution as defined above, the distribution that this protocol recursively runs on is still a product distribution.

\textbf{Analysis of $\phi^0_{v, \gamma, 1/2 - \epsilon}$:} First we should make sure that the probabilities we use to sample $b_x$ and $b_y$ are no greater than 1. When Alice and Bob proceed to sample $b_x$ and $b_y$, we know that $threshold_{\theta, v, w, D}$ returns 0. Therefore $m_x(v,w) \leq \theta_x$ and $m_y(v,w)  \leq \theta_y$. So
\begin{eqnarray*}
\frac{(1/2 - \epsilon)^{m_x(v,w) - \theta_x}(1/2+\epsilon)^{- m_x(v,w) + \theta_x}}{(1/2 - 2\epsilon)^{m_x(v,w) - \theta_x}(1/2+2\epsilon)^{ - m_x(v,w) + \theta_x}} &=& \left(\frac{1/2-\epsilon}{1/2-2\epsilon}\right)^{m_x(v,w) - \theta_x}\left(\frac{1/2+\epsilon}{1/2+2\epsilon}\right)^{\theta_x - m_x(v,w)} \\
&=& \left(\frac{1/4 - \epsilon /2 - 2\epsilon^2}{1/4 + \epsilon /2 - 2\epsilon^2}\right)^{\theta_x - m_x(v,w)}\leq 1.
\end{eqnarray*}
Similarly, we have 
\[
\frac{(1/2 - \epsilon)^{m_y(v,w) - \theta_y}(1/2+\epsilon)^{- m_y(v,w) + \theta_y}}{(1/2 - 2\epsilon)^{m_y(v,w) - \theta_y}(1/2+2\epsilon)^{ - m_y(v,w) + \theta_y}} \leq 1.
\]
The probability that the protocol accepts some $w$ in each round is: 
\begin{eqnarray*}
&&\sum_{w, m(v,w)\leq \theta}(1/2 - 2\epsilon)^{m(v,w)}(1/2 + 2\epsilon)^{\gamma - m(v,w)}\cdot \frac{(1/2 - \epsilon)^{m_x(v,w) - \theta_x}(1/2+\epsilon)^{- m_x(v,w) + \theta_x}}{(1/2 - 2\epsilon)^{m_x(v,w) - \theta_x}(1/2+2\epsilon)^{ - m_x(v,w) + \theta_x}}\\
&&\cdot \frac{(1/2 - \epsilon)^{m_y(v,w) - \theta_y}(1/2+\epsilon)^{- m_y(v,w) + \theta_y}}{(1/2 - 2\epsilon)^{m_y(v,w) - \theta_y}(1/2+2\epsilon)^{ - m_y(v,w) + \theta_y}}\\
&=&\sum_{w, m(v,w)\leq \theta}(1/2 - 2\epsilon)^{m(v,w)}(1/2 + 2\epsilon)^{\gamma - m(v,w)}\cdot \frac{(1/2 - \epsilon)^{m(v,w) - \theta}(1/2+\epsilon)^{- m(v,w) + \theta}}{(1/2 - 2\epsilon)^{m(v,w) - \theta}(1/2+2\epsilon)^{ - m(v,w) + \theta}}\\
&=& \sum_{w, m(v,w)\leq \theta}(1/2 - \epsilon)^{m(v,w)}(1/2 + \epsilon)^{\gamma - m(v,w)}\cdot 
\frac{(1/2-2\epsilon)^{\theta}(1/2 + 2\epsilon)^{\gamma - \theta}}{(1/2-\epsilon)^{\theta}(1/2 + \epsilon)^{\gamma - \theta}}\\
&=& p \cdot \frac{(1/2-2\epsilon)^{\theta}(1/2 + 2\epsilon)^{\gamma - \theta}}{(1/2-\epsilon)^{\theta}(1/2 + \epsilon)^{\gamma - \theta}}\\
&=& p \cdot \frac{(1/2-2\epsilon)^{(1/\epsilon^2)(1/2 - 3\epsilon)}(1/2 + 2\epsilon)^{(1/\epsilon^2)(1/2 +3\epsilon)}}{(1/2-\epsilon)^{(1/\epsilon^2)(1/2 - 3\epsilon)}(1/2 + \epsilon)^{(1/\epsilon^2)(1/2 + 3\epsilon)}}\\
&\geq& 5p.
\end{eqnarray*}
The last inequality comes from the following argument:
\begin{eqnarray*}
&&\lim_{\epsilon \rightarrow 0} \frac{(1/2-2\epsilon)^{(1/\epsilon^2)(1/2 - 3\epsilon)}(1/2 + 2\epsilon)^{(1/\epsilon^2)(1/2 +3\epsilon)}}{(1/2-\epsilon)^{(1/\epsilon^2)(1/2 - 3\epsilon)}(1/2 + \epsilon)^{(1/\epsilon^2)(1/2 + 3\epsilon)}}\\
&=& \lim_{\epsilon \rightarrow 0}\left(1-\frac{12\epsilon^2}{1-4\epsilon^2}\right)^{\frac{1}{2\epsilon^2}- \frac{3}{\epsilon}}\left(1+\frac{2\epsilon}{1+2\epsilon}\right)^{\frac{6}{\epsilon}}\\
&=& \lim_{\epsilon \rightarrow 0}  \exp\left(-\frac{12\epsilon^2}{1-4\epsilon^2}\cdot \left(\frac{1}{2\epsilon^2} - \frac{3}{\epsilon}\right) + \frac{2\epsilon}{1+2\epsilon} \cdot \frac{6}{\epsilon}\right)\\
&=& e^6.\\
\end{eqnarray*}
So there exists a constant $\beta$ such that, when $0 < \epsilon < \beta$, 
\[
\frac{(1/2-2\epsilon)^{(1/\epsilon^2)(1/2 - 3\epsilon)}(1/2 + 2\epsilon)^{(1/\epsilon^2)(1/2 +3\epsilon)}}{(1/2-\epsilon)^{(1/\epsilon^2)(1/2 - 3\epsilon)}(1/2 + \epsilon)^{(1/\epsilon^2)(1/2 + 3\epsilon)}}\geq 5.
\]
Therefore the expected number of rounds is at most $\frac{1}{5p}$. By induction, each call of $\phi_{v,\gamma, 1/2 - 2\epsilon}$ uses at most $\alpha \cdot (2\epsilon)^2 \cdot \gamma$ bits of communication in expectation. Thus the expected number of bits communicated in $\phi^0_{v, \gamma, 1/2 - \epsilon}$ is 
\[
\frac{1}{5p} (\alpha \cdot (2\epsilon)^2 \cdot \gamma + 2 + 8) = \frac{4\alpha}{5p} + \frac{2}{p}.
\]
From the above analysis, we can also see that for a specific node $w$ with $m(v,w) \leq \theta$, the probability that $w$ is sampled and accepted in each round of this protocol is 
\[
(1/2 - \epsilon)^{m(v,w)}(1/2 + \epsilon)^{\gamma - m(v,w)}\cdot 
\frac{(1/2-2\epsilon)^{\theta}(1/2 + 2\epsilon)^{\gamma - \theta}}{(1/2-\epsilon)^{\theta}(1/2 + \epsilon)^{\gamma - \theta}}.
\]
Then since the probability that the protocol ends in each round is
\[
p \cdot \frac{(1/2-2\epsilon)^{\theta}(1/2 + 2\epsilon)^{\gamma - \theta}}{(1/2-\epsilon)^{\theta}(1/2 + \epsilon)^{\gamma - \theta}},
\]
the probability that $w$ with $m(v,w) \leq \theta$ is sampled in this protocol is
\[
\frac{
(1/2 - \epsilon)^{m(v,w)}(1/2 + \epsilon)^{\gamma - m(v,w)}\cdot 
\frac{(1/2-2\epsilon)^{\theta}(1/2 + 2\epsilon)^{\gamma - \theta}}{(1/2-\epsilon)^{\theta}(1/2 + \epsilon)^{\gamma - \theta}}}{p \cdot \frac{(1/2-2\epsilon)^{\theta}(1/2 + 2\epsilon)^{\gamma - \theta}}{(1/2-\epsilon)^{\theta}(1/2 + \epsilon)^{\gamma - \theta}}} = \frac{(1/2 - \epsilon)^{m(v,w)}(1/2 + \epsilon)^{\gamma - m(v,w)}}{p}.
\]

\textbf{Analysis of $\phi^1_{v, \gamma, 1/2 - \epsilon}$:} First we should make sure that the probabilities we use to sample $b_x$ and $b_y$ are no greater than 1. Recall that $t=e^6$. When Alice and Bob proceed to sample $b_x$ and $b_y$, we know that $threshold_{\theta, v, w, D}$ returns 1. Therefore $m_x(v,w) \geq \theta_x$ and $m_y(v,w)  \geq \theta_y$. So
\begin{eqnarray*}
\frac{(1/2-\epsilon)^{m_x(v,w)}(1/2 +\epsilon)^{\gamma/2 -m_x(v,w)}}{t\cdot (\frac{1/2 - \epsilon}{1/2 +\epsilon})^{\theta_x - \theta / 2} \cdot 2^{-\gamma/2}} &\leq& \frac{(1/2 - \epsilon)^{\theta /2}(1/2 + \epsilon)^{\gamma/2 - \theta/2}}{t\cdot2^{-\gamma/2}}\\
&=&\frac{(1-4\epsilon^2)^{\theta / 2} (1+2\epsilon)^{\gamma /2 - \theta}}{e^6} \\
&\leq& \frac{(1+2\epsilon)^{3/\epsilon}}{e^6} \leq 1.\\
\end{eqnarray*}
Similarly, we have
\[
\frac{(1/2-\epsilon)^{m_y(v,w)}(1/2 +\epsilon)^{\gamma/2 -m_y(v,w)}}{t\cdot (\frac{1/2 - \epsilon}{1/2 +\epsilon})^{\theta_y - \theta / 2} \cdot 2^{-\gamma/2}} \leq 1.
\]
The probability that the protocol accepts some $w$ in each round is:
\begin{eqnarray*}
&&\sum_{w,m(v,w)> \theta} 2^{-\gamma}\cdot \frac{(1/2-\epsilon)^{m_x(v,w)}(1/2 +\epsilon)^{\gamma/2 -m_x(v,w)}}{t\cdot (\frac{1/2 - \epsilon}{1/2 +\epsilon})^{\theta_x - \theta / 2} \cdot 2^{-\gamma/2}} \cdot \frac{(1/2-\epsilon)^{m_y(v,w)}(1/2 +\epsilon)^{\gamma/2 -m_y(v,w)}}{t\cdot (\frac{1/2 - \epsilon}{1/2 +\epsilon})^{\theta_y - \theta / 2} \cdot 2^{-\gamma/2}}\\
&=& \sum_{w,m(v,w)> \theta} (1/2-\epsilon)^{m(v,w)} (1/2+\epsilon)^{\gamma - m(v,w)} \cdot \frac{1}{t^2}\\
&=& \frac{1-p}{t^2}.
\end{eqnarray*}
Therefore the expected number of bits communicated in $\phi^1_{v, \gamma, 1/2 - \epsilon}$ is at most
\[
\frac{t^2}{1-p} (2 + 8) = \frac{10t^2}{1-p}.
\]
From the above analysis, we can also see that for a specific node $w$ with $m(v,w) > \theta$, the probability that $w$ is sampled and accepted in each round is 
\[
(1/2-\epsilon)^{m(v,w)} (1/2+\epsilon)^{\gamma - m(v,w)} \cdot \frac{1}{t^2}.
\]
Then since the probability that the protocol ends each round is $\frac{1-p}{t^2}$,
the probability that $w$ with $m(v,w) \leq \theta$ is sampled in this protocol is
\[
\frac{(1/2-\epsilon)^{m(v,w)} (1/2+\epsilon)^{\gamma - m(v,w)} \cdot \frac{1}{t^2}}{\frac{1-p}{t^2}} = \frac{(1/2-\epsilon)^{m(v,w)} (1/2+\epsilon)^{\gamma - m(v,w)}}{1-p}
\]

\textbf{Analysis of $\phi_{v, \gamma, 1/2 - \epsilon}$:} Combining the analysis of $\phi^0_{v, \gamma, 1/2 - \epsilon}$ and $\phi^1_{v, \gamma, 1/2 - \epsilon}$, the expected number of bits communicated in $\phi_{v, \gamma, 1/2 - \epsilon}$ is at most
\[
p \cdot ( \frac{4\alpha\epsilon^2\gamma}{5p} + \frac{2}{p}) + (1-p)\cdot \frac{10t^2}{1-p} = \frac{4\alpha}{5} + 2 + 10t^2 \leq \frac{4\alpha}{5} + \frac{\alpha}{5} = \alpha.
\]
For node $w$ with $m(v,w) > \theta$, the probability that $w$ is sampled is
\[
\frac{(1/2 - \epsilon)^{m(v,w)}(1/2 + \epsilon)^{\gamma - m(v,w)}}{p} \cdot p = (1/2 - \epsilon)^{m(v,w)}(1/2 + \epsilon)^{\gamma - m(v,w)}.
\]
For node $w$ with $m(v,w) \leq \theta$, the probability that $w$ is sampled is 
\[
\frac{(1/2 - \epsilon)^{m(v,w)}(1/2 + \epsilon)^{\gamma - m(v,w)}}{1-p} \cdot (1-p) = (1/2 - \epsilon)^{m(v,w)}(1/2 + \epsilon)^{\gamma - m(v,w)}.
\]
So all the nodes are sampled according to the correct probability distribution.
\qed

\section{Distributional energy cost is equal to external information cost} 
\begin{thm}\label{eclb}
For any protocol $\pi$ over a variable-error binary symmetric channel with feedback and any distribution $\mu$ over inputs, there is a private coin protocol $\phi$ over the noiseless binary channel, such that $ IC_{\mu}^{ext}(\phi) \leq \frac{1}{\ln(2)}EC_{\mu}(\pi) $ and $\phi$ simulates $\pi$. 
\end{thm}
\proof
Protocol $\phi$ is very simple to be constructed from $\pi$. For each transmitted bit $b$ in $\pi$, if the transmitter wants to send $b$ over $BSC_p$ in step $i$, the transmitter in $\phi$ sends $b \oplus N_i$ to the receiver, where $N_i\sim B_p$ is a Bernoulli random variable with probability $p$ of being $1$.  It is clear that $\phi$ simulates $\pi$. 

Now let's analyze the external information cost of $\phi$. By definition and Fact \ref{cr}, 
\[
IC_{\mu}^{ext}(\phi) = I(XY;\Phi) = \sum_{i=1}^{CC(\phi)} I(XY;\Phi_i|\Phi_{<i}).
\]
By Fact \ref{div}, we have
\[
I(XY;\Phi_i |\Phi_{<i}) = \mathbb{E}_{x,y,\phi_{<i}} [D((\Phi_i|X=x,Y=y,\Phi_{<i} = \phi_{<i})\| (\Phi_i|\Phi_{<i} = \phi_{<i}))].
\]
Now fix $\Phi_{<i} = \phi_{<i}$, let $p(x,y) = Pr[(\Phi_i|X=x,Y=y,\Phi_{<i} = \phi_{<i}) = 1]$ and $q = Pr[(\Phi_i|\Phi_{<i} = \phi_{<i}) = 1]$. Then we have,
\[
 \mathbb{E}_{x,y} [D((\Phi_i|X=x,Y=y,\Phi_{<i} = \phi_{<i})\| (\Phi_i|\Phi_{<i} = \phi_{<i}))] =  \mathbb{E}_{x,y}[D(p(x,y)\|q)],
\]
and
\[
\mathbb{E}_{x,y}[p(x,y)] = q.
\]
By the definition of KL-divergence,
\[
\mathbb{E}_{x,y}\left[D(p(x,y)\|q) - D(p(x,y)\| \frac{1}{2})\right] = \mathbb{E}_{x,y}\left[p(x,y)\log(\frac{1}{2q}) + (1-p(x,y))\log(\frac{1}{2(1-q)})\right] = H(q) - 1 \leq 0.
\]
Note that $\pi$ may use private randomness. Let $r$ be the private randomness of the party whose turn it is to speak, and let 
$p(x,y,r)$ be the probability  $Pr[(\Phi_i|X=x,Y=y,R=r,\Phi_{<i} = \phi_{<i}) = 1]$. Then $p(x,y)=\mathbb{E}_{r|x,y} p(x,y,r)$. 
By combining the previous line with Fact \ref{ine} and the convexity of $z\mapsto z^2$  we have, 
\begin{multline*}
 \mathbb{E}_{x,y}[ D(p(x,y)\| q)] \leq
 \mathbb{E}_{x,y}[ D(p(x,y)\| \frac{1}{2})]  \leq  \mathbb{E}_{x,y}\left[\frac{4}{\ln(2)}(p(x,y) - \frac{1}{2})^2\right] \leq \\ 
\mathbb{E}_{x,y,r}\left[\frac{4}{\ln(2)}(p(x,y,r) - \frac{1}{2})^2\right] = \frac{1}{\ln(2)}\mathbb{E}_{x,y}[EC(\Pi_i) | X = x, Y=y, \Pi_{<i} = \phi_{<i} ].
\end{multline*}
To sum up, we get
\begin{eqnarray*}
IC_{\mu}^{ext}(\phi) &=& \sum_{i=1}^{CC(\phi)} I(XY;\Phi_i |\Phi_{<i}) \\
&=& \sum_{i=1}^{CC(\phi)}  \mathbb{E}_{x,y,\phi_{<i}} [D((\Phi_i|X=x,Y=y,\Phi_{<i} = \phi_{<i})\| (\Phi_i|\Phi_{<i} = \phi_{<i}))] \\
&\leq& \frac{1}{\ln(2)}\sum_{i=1}^{CC(\pi)}  \mathbb{E}_{x,y,\phi_{<i}}[EC(\Pi_i) | X = x, Y=y, \Pi_{<i} = \phi_{<i}] \\
&=& \frac{1}{\ln(2)} EC_{\mu}(\pi).
\end{eqnarray*}
\qed

\begin{thm}\label{ecub}
For any protocol $\pi$ over a noiseless channel, any distribution $\mu$ over inputs and any $\epsilon = \frac{1}{2n}, n\in \mathbb{Z}, n >0$, there is a protocol $\phi$ over  a variable-error binary symmetric channel with feedback, such that $EC_{\mu}(\phi) = O(IC_{\mu}^{ext}(\pi) + \epsilon )$ and $\phi$ simulates $\pi$. 
\end{thm}
\proof
Similarly to the proof of Theorem \ref{eclb}, we first express the external information cost of $\pi$ as the sum of the divergence between the true probability and the prior probability. Let $p_{x,y,i,\pi_i} = Pr[(\Pi_i|X = x,Y=y,\Pi_{<i} =\pi_{<i}) = 1]$ and $q_{i,\pi_i} = Pr[(\Pi_i|\Pi_{<i} = \pi_{<i}) = 1]$. Then we have
\begin{eqnarray*}
IC_{\mu}^{ext}(\pi) &=& \sum_{i = 1}^{CC(\pi)}\mathbb{E}_{x,y,\pi_{<i}}[D((\Pi_i|X = x,Y=y,\Pi_{<i} =\pi_{<i})\|(\Pi_i|\Pi_{<i} = \pi_{<i}))]\\
&=& \sum_{i = 1}^{CC(\pi)}\mathbb{E}_{x,y,\pi_{<i}}[D(p_{x,y,i,\pi}\| q_{i,\pi_i})].
\end{eqnarray*}
We are going to construct $\phi$ by simulating $\pi$'s communication bit by bit. For the $i$th transmitted bit, given inputs $x,y$ and the previous transcript $\pi_{<i}$, it is sufficient to prove that the corresponding simulation in $\phi$ uses energy cost at most $O(D(p_{x,y,i,\pi_{<i}} \| q_{i,\pi_{<i}})+\frac{\epsilon}{2^i})$ in expectation, and the receiver can sample a bit from Bernoulli distribution $B_{p_{x,y,i,\pi_{<i}}}$ given prior $q_{i,\pi_{<i}}$. 

Now, we construct the simulation of the $i$th transmitted bit given inputs $x,y$ and the previous transcript $\pi_{<i}$. Since we fix $i,x,y,\pi_{<i}$ here, we will abbreviate $p_{x,y,i,\pi_{<i}}$ and $q_{i,\pi_{<i}}$ as $p$ and $q$. The main framework of the construction has following steps: Let $n_i = n \cdot 2^i$ and $\epsilon_i = \frac{1}{2n_i}$. Alice and Bob agree to do biased random walk on points $0, \frac{1}{2n_i}, \frac{2}{2n_i}, \cdots, \frac{2n_i-1}{2n_i}, 1$, starting at a point closest to $q$. For each step, the transmitter sends one bit over some binary symmetric channel with some chosen crossover probability. They move right for one step if the received bit is 1, and they move left for one step if the the received bit is 0. They stop this random walk whenever they reach 0 or 1, and take the value on the point as the corresponding sampled bit. As $\sum_{i=1}^{CC(\pi)} \epsilon_i \leq \epsilon$, it is sufficient to prove that the energy cost of this communication of $O(D(p\|q)+\epsilon_i)$ and after random walk they reach 1 with probability $p$. Note that setting $\epsilon_i$ in this way is for the case when $\pi$ has finite external information cost but a potentially unbounded communication complexity. Otherwise we can pick $\epsilon_i = \frac{\epsilon}{CC(\pi)}$. 

We need the following lemma as the main technique of our construction. 
\begin{lem}\label{brw}
Suppose Alice and Bob agree to do biased random walk on points $0,1,\ldots,a+b$ via communication over binary symmetric channels, and they start at point $a$. If $a \geq b$, the transmitter only has to send messages with energy cost at most $48$ to make them always end at $a+b$.
\end{lem}

\proof We prove this lemma by induction on $(a+b)^2+b$, showing that the lemma for smaller $(a+b)^2+b$ implies it for larger $(a+b)^2+b$. The basis of this induction proof is the case when $b \leq 12$. If $b \leq 12$, the transmitter only has to send 1 over $BSC_0$ (noiseless channel) for $b$ times. This will take at most $12< 48$ energy cost and they will end at $a+b$. 

If $b > 12$. Let $c = \lfloor \frac{a}{2} \rfloor$. The protocol is as follows:
\begin{Protocol}
\caption{Biased Random Walk}
\begin{enumerate}
\item They first do biased random walk on points $a-c,a-c+1,\cdots, a, \cdots , a+b$ with start point $a$. For each step, the transmitter sends 1 over $BSC_{\frac{1}{2} - \frac{3}{c}}$. They stop this procedure when they reach either $a-c$ or $a+b$, or they have already taken $c^2$ steps. Suppose they stop at point $d$.
\item If they reach $d = a+b$, the protocol ends.
\item If $d < a$, we know that $d \geq a- c$. By induction, they do biased random walk on points $0,...,a$ with start point $d$ to get back to $a$. And then they run this protocol again.
\item If $d=a$, they run this protocol again.
\item If $d > a$, by induction, they do biased random walk on points $1,...,a+b$ with start point $d$ to get to $a+b$ and the protocol ends.  
\end{enumerate}
\end{Protocol}
\\
Let's analyze this protocol. First we calculate the probability that they reach point $a-c$ after the first part of the protocol. This probability is no more than the probability of reaching $a-c$ if we change the stop condition of the first part to stopping only when reaching either $a-c$ or $a+b$. We can calculate the second probability by recursion. For $(\frac{1}{2} + \frac{3}{c})$-biased random walk on points $a-c,...,a+b$ with start point $t$, define $u_{t}$ to be the probability of reaching $a+b$. Then we have $u_{a-c} = 0$, $u_{a+b} = 1$ and $u_t = (\frac{1}{2} + \frac{3}{c})u_{t+1} +  (\frac{1}{2} - \frac{3}{c})u_{t-1}$ for $a-c < t < a+b$. Let $\beta = \frac{\frac{1}{2} - \frac{3}{c}}{\frac{1}{2} + \frac{3}{c}}$, we have 
\[
u_t = \frac{1+ \cdots + \beta^{t -(a-c)-1}}{1+ \cdots +\beta^{b+c - 1}}.
\]
Since $c + 1  = \lfloor \frac{a}{2} \rfloor + 1\geq b / 2$ and $b > 12$, we know $3c \geq b$. Then we have
\[
u_a = \frac{1+ \cdots + \beta^{c- 1}}{1+ \cdots +\beta^{b+c - 1}} \geq \frac{1}{1+\beta^c + \beta^{2c} + \beta^{3c}} > \frac{1}{1+3\beta^c}.
\]
We also have
\[
\beta^c = \left(1 - \frac{\frac{6}{c}}{\frac{1}{2} + \frac{3}{c}}\right)^c \leq \left(1-\frac{6}{c}\right)^c < e^{-6}.
\]
So
\[
u_a > \frac{1}{1+3\beta^c} > \frac{1}{1 + 3e^{-6}}.
\]
Therefore, for the first part of the protocol, the probability of reaching $a-c$ is at most $1 - \frac{1}{1 + 3e^{-6}}$.

Now let's calculate the probability of stopping at point between $a-c$ and $a+b$ after $c^2$ steps of $(\frac{1}{2} + \frac{3}{c})$-biased random walk . For each step, with probability $\frac{1}{2} + \frac{3}{c}$, the coordinate will increase 1, and with probability $\frac{1}{2} - \frac{3}{c}$ the coordinate will decrease 1. If $a-c < d < a+b$, the sum of these values will be less than $b$. By Chernoff bound, the probability that the sum of these values is less than $b$ is no more than
\[
e^{-\frac{2(6c-b)^2}{4c^2}} < e^{-\frac{2(3c)^2}{4c^2}} = e^{-4.5}.
\]
So the probability that $a-c < d < a+b$ is at most $e^{-4.5}$. 

Now we can calculate the expected energy cost of this protocol. Let's assume the expected energy cost of this protocol is $v$. For the first part of the protocol, it takes $4(\frac{1}{2} - \frac{3}{c} - \frac{1}{2})^2\cdot c^2 = 36$ energy cost. If $a-c \leq d <a $, the protocol will spend at most $v + 48$ energy cost after the first part.  If $d = a$, the protocol will spend at most $v$ after the first part. If $a <d < a+b$, the protocol will spend at most $48$ energy. So if $d \neq a+b$, the protocol will spend at most $v + 48$ energy cost after the first part. Using the probability we calculate before, we have
\[
v \leq (e^{-4.5} + 1 - \frac{1}{1 + 3e^{-6}})(v + 48) + 36 \leq (\frac{1}{16} + \frac{1}{16})(v+48) + 36 = \frac{v}{8} + 42.
\]
Therefore $v \leq 48$ as desired. 
\qed

Directly from this lemma, the transmitter can go from point $q$ to point $2^t\cdot q$ with energy cost $O(t)$ by applying the protocol in this lemma $t$ times. 

Let's start the construction. Without loss of generality, let's assume $0 < q \leq \frac{1}{2}$. Notice that we ignore the case when $q= 0$. Because if $q=0$, $p$ must be 0 and the receiver can sample one bit from $B_p$ without any communication. Now we assume $2n_iq$ is an integer and we will consider the case that $2n_iq$ is not an integer later in the proof. The general protocol of sampling one bit from Bernoulli distribution $B_p$ given prior $q$ is as Protocol 6.

\begin{Protocol}
\caption{General Protocol}
\begin{enumerate}
\item Let $n_i = n \cdot 2^i$ and $\epsilon_i = \frac{1}{2n_i}$. Alice and Bob agree to do some biased random walk on points $0 ,\frac{1}{2n_i}, \cdots, \frac{1}{2}$ with start point $q$. 
\item If they end at point 0, then 0 is the sampled bit.
\item If they end at point $\frac{1}{2}$, the transmitter will send one more bit over some binary symmetric channel, and the received bit will be taken as the sampled bit. 
\end{enumerate}
\end{Protocol}

The energy cost we are going to use when $2qn_i$ is an integer is $O(D(p\| q))$. We use different lower bounds of $D(p\|q)$ for different values of $p$ and $q$. In all the cases, Alice and Bob will follow the general protocol. The only difference is that for different cases, the transmitter will choose different biases for biased random walk. The detailed differences are shown in Protocol 7.

\begin{Protocol}
\caption{Detailed Protocols in cases}
\begin{enumerate}
\item If $0 \leq p \leq 2q$, the transmitter will first send 1's over $BSC_{\frac{1}{2}}$ until they reach point $\frac{1}{2}$ or $0$. Suppose they reach $\frac{1}{2}$, the transmitter will send 1 over $BSC_{1 - \frac{p}{2q}}$ if $p \geq q$, and send 0 over $BSC_{\frac{p}{2q}}$ if $p < q$. 
\item If $ 2q < p < 0.02$, $q < 0.01$, the transmitter will first send 1's over $BSC_{\frac{1}{2}}$ until they reach point $\frac{\lfloor \frac{2n_iq}{p} \rfloor}{2n_i}$ or 0. If they reach $\frac{\lfloor \frac{2n_iq}{p} \rfloor}{2n_i}$, the transmitter will use the protocol in Lemma \ref{brw} $O(\log(\frac{p}{q}))$ times to arrive $\frac{1}{2}$. Finally the transmitter will send 1 over $BSC_{1-\frac{p\lfloor \frac{2n_iq}{p} \rfloor}{2n_iq} }$. 

\item Otherwise, the transmitter will use the protocol in Lemma \ref{brw} $O(\log(\frac{1}{q}))$ times to arrive $\frac{1}{2}$. Then the transmitter will send 1 over $BSC_{1-p}$ if $p \geq \frac{1}{2}$, and send 0 over $BSC_{p}$ if $p \leq \frac{1}{2}$. 
\end{enumerate}
\end{Protocol}

To analyze these protocols, we need the following simple lemma:
\begin{lem}\label{ubrw}
Suppose Alice and Bob agree to do unbiased random walk on points $0,1,...,a+b$ via communication over $BSC_{\frac{1}{2}}$, and they start at point $a$. Then the probability that they end at $a+b$ is $\frac{a}{a+b}$. 
\end{lem}
\proof For unbiased random walk on points $0,...,a+b$ with start point $t$, define $u_{t}$ to be the probability of reaching $a+b$. Then we have $u_0 = 0$, $u_{a+b} = 1$ and $u_t = \frac{1}{2}(u_{t-1} + u_{t+1})$ for $0<t<a+b$. Solve this we get $u_t = \frac{t}{a+b}$ and thus $u_{a} = \frac{a}{a+b}$. 
\qed

Now we are going to show in cases that the detailed protocols sample a bit from Bernoulli distribution $B_p$ and use energy cost $O(D(p\|q))$ in expectation. Notice that although the last 2 cases use the same protocol, as we use different lower bounds of $D(p\|q)$ in these 2 cases, we have to analyze them separately.
\begin{enumerate}
\item $0 \leq p \leq 2q$: By Lemma \ref{ubrw}, after unbiased random walk, the probability that they reach $\frac{1}{2}$ is $2q$. So the probability that the sample bit is 1 is $2q \times \frac{p}{2q} = p$. By Fact \ref{ine}, $D(p\|q) = \Omega(\frac{(p-q)^2}{q})$. The energy cost of the protocol only comes from the last bit, which equals to $2q \times 4(\frac{p}{2q} -\frac{1}{2})^2 = O\left(\frac{(p-q)^2}{q}\right) = O(D(p\|q))$. 
\item $ 2q < p < 0.02$, $q < 0.01$: By Lemma \ref{ubrw}, after unbiased random walk, the probability that they reach $\frac{\lfloor \frac{2n_iq}{p} \rfloor}{2n_i}$ is $\frac{2n_iq}{\lfloor \frac{2n_iq}{p} \rfloor}$. So the probability that they get sample bit 1 is $\frac{2n_iq}{\lfloor \frac{2n_iq}{p} \rfloor} \times \frac{p\lfloor \frac{2n_iq}{p} \rfloor}{2n_iq} = p$.

Now let's give the lower bound of $D(p\|q)$ in this case.
\begin{itemize}
\item  If $p >3q$, then $p\log \frac{p}{q} \geq p\log (3)$ and 
\[
|(1-p)\log \frac{1-p}{1-q}|  = (1-p)\log \frac{1-q}{1-p} = (1-p )\log(1 + \frac{p-q}{1-p}) < p - q \leq p.
\]
So 
\[
D(p\| q) = p\log \frac{p}{q}  + (1-p)\log \frac{1-q}{1-p} \geq p\log \frac{p}{q} - p \geq (1-1/\log(3))p\log \frac{p}{q} = \Omega\left(p\log \frac{p}{q}\right).
\]
\item If $2q < p \leq 3q$, by Fact \ref{ine}, $D(p\| q) = \Omega(p)$ and $p\log \frac{p}{q} = O(p)$, so $D(p\| q) =  \Omega(p\log \frac{p}{q})$. 
\end{itemize}
So in this case $D(p\|q) = \Omega(p\log\frac{p}{q})$. The energy cost of this protocol comes from the biased random walk which has energy cost $O(\log \frac{p}{q})$ and the last bit which has energy cost at most 1. As $2n_iq \geq 1$, the probability that the transmitter has to do biased random walk and to send the last bit is 
\[
\frac{2n_iq}{\lfloor \frac{2n_iq}{p} \rfloor}  < \frac{2n_iq}{\frac{2n_iq}{p}  - 1} = p \times \frac{2n_iq}{2n_iq - p} = p \frac{1}{1- \frac{p}{2n_iq}}  < p \frac{1}{1-p} = O(p).
\]
Therefore the total energy cost is at most
\[
O(p (\log \frac{p}{q} + 1)) = O(p\log\frac{p}{q}) = O(D(p\|q)).
\]

\item $p > 2q$ and $q \geq 0.01$: From the protocol, we know that they will always arrive $\frac{1}{2}$ after biased random walk. Then after the last step, the probability that they get 1 is $p$. For the lower bound of $D(p\|q)$, since $p-q > q \geq 0.01$, by Fact \ref{ine}, $D(p\|q) = \Omega((p-q)^2) = \Omega(1)$. The energy cost of the protocol is 
\[
O(\log \frac{1}{q} + 1) = O(1) = O(D(p\| q)).
\]

\item $p \geq 0.02, q < 0.01$: Similarly as the previous case, the probability that the sampled bit is 1 is $p$. Now we give the lower bound of $D(p\|q)$.
\begin{itemize}
\item If $0.005<q < 0.01$, then 
\[
D(p\| q) = \Omega(1) = \Omega(\log \frac{1}{q}).
\]
\item If $q \leq 0.005$, then
\[
D(p\| q) = p\log \frac{1}{q} + (1-p)\log \frac{1}{1-q} - H(p) \geq p\log \frac{1}{q} - H(p).
\]
Now let's consider $H(p) /p$. If $p > 1/5$, then $H(p) / p < 1 / (1/5) = 5$. If $0.02< p \leq 1/5$,
\[
H(p)/ p = \log\frac{1}{p} + \frac{1-p}{p}\log(1 + \frac{p}{1-p}) \leq \log \frac{1}{p} + \frac{1-p}{p} \cdot \frac{p}{1-p} < \log(50) + 1 = \log (100).
\]
Therefore, for all $p \geq 0.02$, $H(p) / p < \log(100)$. So
\[
\frac{p \log \frac{1}{q}}{H(p)} \geq \frac{\log(200)}{\frac{H(p)}{p}} > \frac{\log(200)}{\log(100)}.
\]
Thus
\[
D(p\| q) \geq p\log \frac{1}{q} - H(p) \geq \left(1 - \frac{\log(100)}{\log(200)}\right) p\log \frac{1}{q} = \Omega\left(\log \frac{1}{q}\right).
\]
\end{itemize}
So in this case, $D(p\|q) = \Omega(\log \frac{1}{q})$. The energy cost of the protocol is 
\[
O(\log \frac{1}{q} + 1) = O(\log \frac{1}{q}) = O(D(p\| q)).
\]
\end{enumerate}

After analyzing these four cases, we have shown that when $2n_iq$ is an integer, our protocol can make the receiver sample a bit from Bernoulli distribution $B_p$ and spends energy cost $O(D(p\| q))$. For the case when $2n_iq$ is not an integer, we can pick $q' = \frac{\lceil 2n_iq \rceil }{2n_i}$ and run the above protocol with prior $q'$. Then the receiver can still sample from Bernoulli distribution $B_p$, and the protocol has cost $O(D(p\| q'))$. Since we have
\[
D(p\|q') - D(p\|q) = p\log \frac{q}{q'} + (1-p)\log \frac{1-q}{1-q'} \leq (1-p)\log\left(1+\frac{q' -q}{1-q'}\right) \leq (1-p)\cdot \frac{q-q'}{1-q'} \leq 1\cdot \frac{\epsilon_i}{0.5} = 2\epsilon_i,
\]
the energy cost is at most
\[
O(D(p\|q')) = O(D(p\|q) + \epsilon_i)
\]
as desired.
\qed



\end{spacing}

\bibliographystyle{alpha}
\bibliography{refs}
 
\end{document}